\documentclass{epl}

\title{Fragile vs strong liquids: a saddles ruled scenario}

\author{Andrea Cavagna\inst{1}}
\institute{
  \inst{1} Department of Physics and Astronomy, The University, Oxford Road, 
M13 9PL, Manchester, UK \\
}

\begin{document}

\maketitle

\begin{abstract}
In the context of the energy landscape description 
of supercooled liquids, we propose an explanation for
the different behaviour of fragile and strong liquids. 
Above Goldstein's temperature $T_x$, diffusion 
is interpreted as a motion in the phase space among 
saddles of the potential energy.
Two mechanisms of diffusion then arise: mechanism $A$ takes 
place when the system overcomes potential energy barriers 
along stable uphill directions, while mechanism $B$ consists 
in finding unstable downhill directions out of a saddle.
Depending on the mutual efficiency of $A$ and $B$,
the usual classification of liquids in fragile and 
strong is recovered. 
Moreover, this scenario naturally predicts the possibility 
of a fragile-to-strong crossover when lowering the temperature. 
\end{abstract}

After the seminal paper by Goldstein in 1969 \cite{gold}, it has become
customary to think of the dynamical evolution of supercooled 
liquids in terms of motion of the state point of the system upon 
its rugged potential energy surface \cite{land}. 
More precisely, at low temperatures, but above the glass transition, 
the diffusion of a system at equilibrium can be interpreted as the 
result of two different processes: the thermal relaxation into 
basins defined by the many minima of the potential energy 
({\it intra}-basin relaxation) and the hopping from basin to basin 
by crossing potential energy barriers ({\it inter}-basin relaxation)
\cite{gold}.

Crucial condition for this description to be correct is that the 
two relaxation times $\tau_{intra}$ and $\tau_{inter}$ are well 
separated, that is $\tau_{inter}\gg\tau_{intra}$:
if these time scales are of the same order, it is not
sensible to discriminate between the thermalization inside 
a minimum and the hopping among different minima.
Indeed, the very requirement of the separation between 
these two time scales led Goldstein in \cite{gold} to define and 
estimate a crossover temperature $T_x$, above which this 
hopping-relaxation description is no longer valid, since 
$\tau_{intra}$ and $\tau_{inter}$ are not well separated.
Below $T_x$, on the other hand, crossing of potential energy barriers
by thermal activation becomes the primary mechanism of diffusion.
The Goldstein temperature $T_x$ is in general higher than
the glass transition temperature $T_g$: 
in the interval $T_g < T < T_x$ a supercooled liquid is still 
in equilibrium on experimental times and can be described 
in terms of relaxation into basins and thermally activated hopping 
among them.

Some authors \cite{angell1,sokolov} have subsequently interpreted 
$T_x$ as the temperature below which the ideal Mode Coupling Theory (MCT) 
\cite{MCT} breaks down. This is because MCT is considered unsuitable for 
describing activated dynamics, so that it prescribes a spurious 
dynamical transition at the point where actually barriers crossing 
becomes a dominant process. According to \cite{angell1,sokolov} the temperature 
$T_c$ where MCT locates this transition must therefore be identified 
with $T_x$ .

A useful realization of Goldstein's scenario has been
introduced by Stillinger and Weber \cite{sw}: the phase space is 
partitioned into basins of attraction of the potential energy minima 
and each dynamical configuration is mapped onto the relative minimum, 
giving rise to a purely {\it inter}-basins dynamics. By comparing 
this pseudo-dynamics with the original one
it is possible to directly verify the validity of Goldstein's 
decomposition into {\it intra} and {\it inter}-basins relaxation and
estimate $T_x$ \cite{sri1}. Computationally, the Stillinger-Weber 
scheme is realized by steepest-descent quenching of configurations 
generated by molecular dynamics simulation, and it proves 
remarkably useful in describing deeply supercooled liquids.

The purpose of this Letter is to generalize  
Goldstein's ideas by introducing an energy landscape description of
supercooled liquids which is valid also above $T_x$, and, in so doing,
to reach a better understanding of the physical difference between 
fragile and strong liquids \cite{angell1}. 
Within the Gibbs-DiMarzio entropic theory \cite{gibbs},
it is in fact possible to check {\it a posteriori} that fragile 
liquids must have a Kauzmann temperature $T_K$ \cite{kauzmann}  
close to the glass transition temperature $T_g$, while for strong 
liquids it must be $T_K\sim 0$. 
However, this explanation of the difference between fragile and
strong liquids has the drawback to involve
the Kauzmann temperature $T_K$, whose existence is not universally
accepted. Moreover, the fragile and strong phenomenology occurs at 
temperatures higher than $T_K$ and $T_g$, so that an interpretation
based only on these two quantities seems unsatisfactory.

At high temperatures Goldstein's description breaks down because 
there is no clear separation between vibration inside a minimum 
and hopping among different minima. 
In order to find a new description we have therefore to start from 
this key observation:
{\it the hopping-relaxation scenario breaks down above $T_x$ 
because the system is no longer spending most part of the
time vibrating around minima}. This happens because 
by raising the temperature, and thus the average potential energy, the 
system starts spending most part of the time in regions of the landscape 
where minima are rare. On the other hand, due to the great complexity 
of the potential energy landscape, other stationary points 
exist and are more numerous at such higher energy levels: 
these objects will not be minima in general, but saddles with 
some unstable directions. 
A generalization of Goldstein's description above $T_x$ must 
therefore deal with saddles. 
Before giving a better specification of these objects, 
let us note that the relevance of saddles 
for the dynamics of glassy systems has been already recognized 
in the past in the context of mean-field spin-glasses 
\cite{laloux,noi,vira}.

Saddles are stationary points of the potential energy
with an arbitrary number $K$ of negative eigenvalues
of the Hessian, i.e. of unstable directions.
The {\it index} $K$ can take any integer value from $K=0$ (minima) 
to $K=D$ (maxima), where $D$ is the dimension of the
phase space (for simple liquids $D=N d$, where $N$ is the number of
particles and $d$ the space dimension). The index density is defined
as $k=K/D$, and the potential energy density as $u=U/N$, where $U$
is the extensive potential energy.
It is useful to introduce the average index density $k$ of the
saddles located at a given potential energy level $u$.
An important feature of this function $k(u)$ is that it is
in general an increasing function of $u$: the higher the energy, 
the larger (on average) the number of negative eigenvalues of 
a saddle at that particular energy. This fact has been explicitly 
proved for a Lennard-Jones liquid in \cite{sad}, where $k(u)$ 
has been found to be a monotonously increasing function.

When considering generic stationary points, it can be introduced a 
generalized notion of potential energy barrier, defined as the extensive 
energy difference $\Delta U$ between a saddle of index $K$ and a higher, 
but adjacent, saddle of index $K+1$. For $K=0$ this is the standard 
definition of energy barrier between a minimum and a simple saddle. 
Provided that we know the average index as a function of the potential 
energy, $k(u)$, we can give a rough estimate of the average barrier $\Delta U$ 
at energy $u$ by evaluating the energy difference between saddles with 
index difference $\Delta K=1$ (see also \cite{laloux}). We have
\begin{equation}
1=\Delta K \sim k'(u) \, \Delta U \, d  \ ,
\end{equation}
and therefore
\begin{equation}
\Delta U(u) \sim [\, k'(u)\, d \,]^{-1} \ .
\label{barra}
\end{equation}
This estimate states that the potential energy barriers between 
stationary points at energy $u$ change according to the change 
in slope of the key function $k(u)$. 
Note that if there is an energy $u_0$ such that 
$k'(u_0)=0$, then barriers diverge at $u_0$.
This happens for example in the $p$-spin spherical spin-glass 
model: here both $k(u)$ and $k'(u)$ vanishes at the so-called 
threshold energy and a purely dynamical transition occurs 
(see, for example, \cite{pspin}). This is in agreement with the 
mean-field nature of that model, which requires the barriers among 
minima to be infinite.
On the other hand, in the Lennard-Jones case $k(u)$ has been found to 
be a linear function, always having nonzero slope \cite{sad}: this
implies that in such a system barriers among saddles are finite and
independent of the energy level.

A generalization of Goldstein's scenario may now be pursued by making
the following hypothesis:
the diffusion of a supercooled liquid can be described as a motion of
the system in the phase space among the neighbourhoods of saddles of 
the potential energy.
In order to give substance to this statement it is important to
specify how to define the neighbourhood of a saddle. This can be
done by considering the effective potential $W\equiv(\vec \nabla U)^2$, 
i.e. the modulus square of the force: all the saddles of the original 
potential energy $U$ are absolute minima of this effective potential $W$ 
\cite{sad,ruocco}. 
At low temperatures the average potential energy of the system 
is small enough for the index $k$ to be on average zero: 
the system stays mainly around minima and  
Goldstein's description is recovered. Indeed, below $T_x$ a quench of a
dynamical configuration on the $W$ surface 
will drain, on average, to a minimum of $U$, and
the present scheme reduces to the Goldstein/Stillinger-Weber 
scenario \cite{nota}. 
However, if the temperature is larger than $T_x$, a quench 
will typically lead to a saddle of $U$, meaning that Goldstein's 
scenario breaks down.

Once assumed that diffusion can be described as a motion among 
the neighbourhoods of different saddles, we can identify two 
mechanisms for this motion to take place \cite{wolf}:

$\bullet \;$ {\bf Mechanism $\bf A \ \cdot \ $}
It consists in the crossing of potential energy barriers:
the system, initially in the neighbourhood of a saddle,
takes an uphill stable direction to reach a saddle with higher energy, 
eventually going downhill to a final saddle at roughly the same 
initial potential energy. Note that, even though we are
considering global stationary points of the {\it whole} potential 
energy, a transition over a potential energy barrier will 
be {\it local} in real space, 
in that the rearrangement process will involve
only a finite number of particles in the system \cite{gold}, with 
all the other particles acting as thermal bath. In this respect,
mechanism $A$ is compatible with both a canonical and a microcanonical 
description of the system.
Note that potential energy is stored at the crossing point. 
At low enough temperatures mechanism $A$ is driven by activation 
and its efficiency directly depends on the temperature through the 
Arrhenius transition probability 
\begin{equation}
P_A\sim \exp(-\Delta U/\kappa T) \ ,
\end{equation}
where $\kappa$ is the Boltzmann constant and $\Delta U$ is the 
potential energy barrier, which can be estimated using the
function $k(u)$. Of course, the efficiency of mechanism $A$ decreases 
by decreasing the temperature. Were this the only mechanism of
diffusion (as in a one-dimensional phase space), a knowledge of the 
exact form of the barriers as a function of the energy would be 
sufficient to predict the behaviour of the system.

$\bullet \;$ {\bf Mechanism $\bf B \ \cdot \ $} 
It exploits the fact that saddles have in general a non-zero
index $K$: in this case the system can find an unstable 
downhill direction, which brings it out of the basin
of the initial saddle. No activation is needed for this to happen.
The system arrives at a lower potential energy level with an excess of 
kinetic energy, which is expended by climbing up again to a new saddle 
at the same initial energy. This mechanism takes place when a given 
cluster of particles suddenly finds a local rearrangement (an unstable 
direction), which sharply decreases their potential energy without any
need of crossing a barrier. 
The extra kinetic energy acquired by the region 
is eventually dissipated by interacting with 
the rest of the system.
Note that, unlike for mechanism $A$, in this case kinetic energy is stored
at the crossing point. 
Mechanism $B$ is the true signature of the multidimensionality 
of the phase space (as opposed to the typical one-dimensional 
picture of barriers hopping) and its nature is entropic, not energetic,
with no direct dependence on the temperature. 
However, there is a crucial {\it indirect} dependence of 
mechanism $B$ on the temperature, due to the fact that 
the average index $k$ (ruling the efficiency of this mechanism) 
is an increasing function of the potential energy of the system,
which is in turn an increasing function of the temperature $T$. 
For example, at very low temperatures the energy of the system is 
so small that only minima are visited on average, and of course mechanism
$B$ is frozen.

Summarizing, the efficiency of both mechanisms decreases when 
the temperature is decreased. A reasonable assumption is 
that, at a given temperature $T$, diffusion is ruled by the most 
efficient of these two competing mechanisms. Therefore we must {\it compare} 
the efficiency of $A$ and $B$ as a function of the temperature,
in order to understand which one of them drives the slowing down 
of the system.

As a first step in this direction we introduce the {\it threshold} 
energy $u_{th}$, defined by the following relation:
\begin{equation}
k(u_{th})\equiv 0  \ .
\label{soglia}
\end{equation}
Below $u_{th}$ unstable saddles become very rare and minima 
dominate, such that mechanism $B$ cannot work. 
Note that, according to relation (\ref{barra}), energy barriers
between threshold minima can be estimated from the slope
of $k(u)$ at the threshold. As already remarked, this slope is 
nonzero in a Lennard-Jones system \cite{sad}, giving finite barriers
among threshold minima. Furthermore, in the Lennard-Jones case it has been 
observed that the threshold energy lies well above the 
energy of the lowest glassy minima found in the system 
\cite{sad}.

A key feature of the threshold energy $u_{th}$ is that it 
allows us to define a critical temperature for mechanism $B$. 
To this aim we have to note that $u_{th}$ is the energy density 
of the threshold minima {\it without considering the vibrational 
contribution}. 
The {\it total} equilibrium average potential energy density 
of a system vibrating around a generic minimum with energy $u_m$, 
can be estimated as
\begin{equation}
u(T)=u_m + \frac{3}{2} \kappa T \ ,
\end{equation}
where $3/2 \kappa T$ is the vibrational contribution within the 
harmonic approximation.
In this way we can {\it define} a critical temperature $T_B$ for 
mechanism $B$ via the following relation:
\begin{equation}
u_{eq}(T_B)-\frac{3}{2} \kappa T_B \equiv u_{th}  \ ,
\label{TB}
\end{equation}
where $u_{eq}(T)$ is the equilibrium average potential energy density 
of the system.
Thus, $T_B$ is the temperature below which the system vibrates 
predominantly around minima, rather than saddles, and mechanism $B$ 
is frozen.
In other words, above $T_B$ the typical saddle has got an extensive number of 
unstable directions, that is $K = O(N)$, and mechanism $B$ 
is highly efficient. At $T_B$ an entropic bottleneck is created,
for $K$ is no longer extensive and the time needed to find
an escape direction from a saddle diverges. 
It is important to understand the following point: at low temperatures
mechanism $A$ may be very slow, but in principle it is always available 
thanks to thermal activation. 
On the contrary, there is no activated regime for mechanism $B$:
when the typical value of $k$ is zero, the system cannot 
borrow an extra direction to escape a minimum. For this reason, 
mechanism $B$ must freeze much more sharply than $A$, passing from 
an efficient phase ($T>T_B$), to a frozen one ($T<T_B$), with no 
activated intermezzo.

In order to compare the two mechanisms we have to ask: What is the
efficiency of mechanism $A$ when mechanism $B$ dies out, that is at $T_B$ ?
To answer this question we must compare the size of the potential
energy barriers at $T_B$ to the amount of thermal energy available 
to activation at this temperature, i.e. $\kappa T_B$. 
The barriers at $T_B$ are given by the potential energy difference 
between threshold minima and simple saddles, 
and their size can be estimated via equation (\ref{barra}) as 
\begin{equation}
\Delta U(u_{th})\sim [\, k'(u_{th})\, d \,]^{-1} \ .
\end{equation}
As already noted, in a Lennard-Jones system $\Delta U(u_{th})$ is
finite \cite{sad}.
We have no {\it a priori} way to know which one of the two quantities,
$\kappa T_B$ or $\Delta U(u_{th})$, will be the largest one.
For the sake of simplicity we will consider the two extremes
cases where one quantity is much larger than the other, classifying 
liquids into two groups:

$\bullet$ {\bf Class I} - The first class of systems is defined by
the relation
\begin{equation}
\Delta U(u_{th}) \ll  \kappa T_B \ .
\label{I}
\end{equation}
The potential energy barriers at $T_B$ are very small, relatively
to the thermal energy, so that mechanism $A$ is still very efficient 
at the temperature where mechanism $B$ freezes: activation is not even 
needed to overcome energy barriers at $T_B$, for too large is the kinetic 
energy of the particles. 
When the temperature is decreased further below $T_B$
mechanism $A$ (the only still available) slows down, and eventually 
Goldstein's temperature $T_x < T_B$ is reached, where 
$\Delta U \sim \kappa T$: in order to pass from minimum to minimum
thermal activation is now needed to cross potential
energy barriers, whereas $B$ is completely unavailable.
In the simplest case where potential energy barriers do not strongly 
depend on $u$, we expect the relaxation time to increase  
according to the Arrhenius law, until the glass transition $T_g$ is 
eventually reached at lower temperatures. For Class I systems thus, 
the slowing down is entirely driven by the slowing down of mechanism $A$, 
i.e. by thermal activation.

$\bullet$ {\bf Class II} - For the second class we have
\begin{equation}
\Delta U(u_{th}) \gg  \kappa T_B \ .
\label{II}
\end{equation}
In this case a very different behaviour may be expected:
potential energy barriers are very large at $T_B$
and therefore {\it mechanism $A$ is already very slow at the  
temperature where B becomes unavailable}. 
Above $T_B$ the system has no 
need to overcome potential energy barriers, because it can use a faster, 
non-thermally activated mechanism of diffusion, that is $B$. 
However, by decreasing the temperature $T_B$ is eventually reached 
and mechanism $B$ suddenly freezes.
Thus, when at $T_B$ for the first time potential energy barriers 
{\it must} be overcome, because no other mechanism of diffusion is 
available, these are already very large compared to $\kappa T_B$. 
In this case, therefore, there must be a sharp increase in the 
relaxation time, entirely driven by the slowing down of mechanism $B$. 
Note that in this case $T_x=T_B$,
because only below $T_B$ activated barriers crossing remains
as the only mechanism of diffusion.
Finally, if barriers are indeed very large at $T_B$, the glass 
transition must occur close to $T_B$ and activation must 
play little role in it.

The smooth, uniform slowing down according to the Arrhenius 
law described for class I corresponds to the {\it strong} liquid 
behaviour, while class II exhibits a sharp dependence of the 
relaxation time on the temperature which is typical of {\it fragile} 
systems. Note that no energy scale, or other dimensional parameter, 
is involved in mechanism $B$, which has a purely geometric nature. 
This fact suggests that the increase of the relaxation time caused 
by the slowing down of $B$ in the fragile case should be well reproduced 
by a power law. This point certainly deserves further investigation.

The validity of the proposed scenario can by directly 
tested  through equations (\ref{barra}), (\ref{soglia}), 
(\ref{TB}), (\ref{I}) and (\ref{II}), provided that the
the average index $k(u)$ as a function of the energy is 
known. This function can be easily computed numerically
in models systems of liquids by sampling saddles of the
potential energy. This program  has been explicitly carried out 
in \cite{sad} for a Lennard-Jones liquid, where it has been found
\begin{equation}
\Delta U(u_{th}) \sim 10 \,\, \kappa T_B \ .
\label{ecco}
\end{equation}
According to the classification given above, relation (\ref{ecco}) is
typical of Class II-fragile systems. This conclusion is in agreement 
with the fragile nature of Lennard-Jones liquids. Furthermore, in 
\cite{sad} it has been proved that $T_B$ coincides with the
MCT critical temperature $T_c$. According to \cite{angell1,sokolov}
this implies $T_B\sim T_x$, in further agreement with our description
for Class II systems.
These facts strongly support the validity of the present description,
at least for the fragile case. 
The same kind of investigation should be performed
in different systems, to see whether equation (\ref{II}) is indeed 
a key feature of fragile liquids.

There is an important phenomenon, whose existence can be 
predicted as a simple corollary of our scenario. 
Consider Class II, but suppose that barriers are 
not very large at $T_B$.
According to what stated above, at $T_B$ the mechanism of diffusion 
switches from $B$ to $A$, and if barriers are now not {\it too} large
here, mechanism $A$ will not be completely unavailable yet. 
For example, we can imagine that the viscosity prescribed by activation
at $T_B$ is 
\begin{equation}
\eta_A(T_B)\sim \exp[\Delta U(u_{th})/\kappa T_B] \sim 10^8 P \ ,
\end{equation}
quite far from the value $\eta\sim 10^{13} P$ defining the glass 
transition. 
What happens is then the following: 
Approaching $T_B$ there is a very sharp jump of $\eta$ (driven by the 
slowing down of $B$) up to the value $\eta\sim 10^8 P$. 
At $T_B$ mechanism $B$ is outstaged by the more efficient 
$A$ and the viscosity starts increasing in a strong, Arrhenius-like 
way according to mechanism $A$, until eventually the glass transition 
occurs at lower temperatures, when $\eta\sim 10^{13} P$. 
A system of this kind must therefore display a fragile high temperature 
phase  for $T>T_B$ and a strong (Arrhenius) low temperature phase for 
$T_g<T<T_B$, that is a {\it fragile-to-strong crossover}. 
Note that the opposite strong-to-fragile transition cannot happen: 
if barriers are small at $T_B$ the system cannot switch back to 
mechanism $B$ at lower temperatures, because there is no activated 
regime for $B$.

Remarkably, there are some recent evidences that a fragile-to-strong 
crossover as the one described above is actually present in
some systems. Preliminary experimental observations 
indicate that this phenomenon occurs in supercooled water 
\cite{acqua}, while it has been quite undoubtedly observed a 
fragile-to-strong crossover in viscous silica, both from 
extrapolations of experimental data \cite{hess} and from 
molecular dynamics simulations \cite{kob}. These are further 
positive tests for the validity of the scenario proposed in 
this Letter.

To conclude, we briefly compare our description with
the Instantaneous Normal Mode (INM) analysis of supercooled 
liquids \cite{stratt}. A fruitful idea in the INM context 
is to relate diffusion to the average fraction of negative
eigenvalues of the potential energy Hessian \cite{keyes}: 
instantaneous unstable modes are 
interpreted as the contribution of regions with negative 
curvature sampled by the system in crossing potential energy barriers. 
In our description unstable directions contribute to 
diffusion only when activation is {\it not} used, that is when 
mechanism $B$ (rather than $A$) rules the dynamics. 
Therefore, one may conclude that the approach of 
\cite{keyes} works so well exactly because above $T_x$ 
activated processes are not relevant. It could be potentially 
very useful to reconsider the role of INM unstable modes 
in the light of our scenario \cite{stratt2}.

\acknowledgments
The author thanks A. Bray, K. Broderix, F. Colaiori, T. Keyes, M.A. Moore, 
G. Parisi, S. Sastry. F. Sciortino, D. Sherrington, F. Thalmann and 
G. Viliani for useful discussions. 
It is a pleasure to acknowledge the collaboration of J.P. Garrahan and 
I. Giardina in an early stage of the work.
This work was supported by EPSRC under grant GR/L97698.


\begin{thebibliography}{}
\bibitem{nota}
It is important to note that the {\it local} minima of $W$ 
do not correspond to zeros of the real force and therefore this 
method does not partition the entire phase space into basins of 
the saddles. It is therefore not correct to define as {\it inherent
saddles} the stationary points found by this method, as done in
\cite{ruocco}.
\end{thebibliography}
\end{document}